\documentclass[12pt]{article}

\usepackage{amssymb,amsmath}
\usepackage{graphicx}
\usepackage{subfigure}

\usepackage{cite}

\newcommand{\be}{\begin{equation}}
\newcommand{\ee}{\end{equation}}

\newcommand{\dlt}{\delta}

\newcommand{\bt}{\beta}

\newcommand{\al}{\alpha}

\newcommand{\gm}{\gamma}
\newcommand{\om}{\omega}

\newcommand{\Gm}{\Gamma}

\newcommand{\lgl}{\langle}
\newcommand{\rgl}{\rangle}

\begin{document}

\begin{center}

{\Large{\bf Optimal conditions for magnetization reversal of nanocluster assemblies 
with random properties}  \\ [5mm]

P.V. Kharebov$^1$, V.K. Henner$^{1,2}$, and V.I. Yukalov$^{3*}$ } \\ [3mm]

$^1$Department of Physics, Perm State University, Perm 614990, Russia \\ [3mm]

$^2$Department of Physics, University of Louisville, Louisville,
Kentucky 40292, USA \\ [3mm]

$^3$Bogolubov Laboratory of Theoretical Physics,  \\
Joint Institute for Nuclear Research, Dubna 141980, Russia

\end{center}

\vskip 2cm

\begin{abstract}

Magnetization dynamics in the system of magnetic nanoclusters with randomly
distributed properties are studied by means of computer simulations. The main
attention is paid to the possibility of coherent magnetization reversal from
a strongly nonequilibrium state with a mean cluster magnetization directed
opposite to an external magnetic field. Magnetic nanoclusters are known to be
characterized by large magnetic anisotropy and strong dipole interactions. It
is also impossible to produce a number of nanoclusters with identical properties.
As a result, any realistic system of nanoclusters is composed of the clusters
with randomly varying anisotropies, effective spins, and dipole interactions.
Despite this randomness, it is possible to find conditions when the cluster
spins move coherently and display fast magnetization reversal due to the
feedback action of resonator. By analyzing the influence of different cluster
parameters, we find their optimal values providing fast magnetization reversal.

\end{abstract}

\vskip 1cm


\vskip 2mm


\vskip 1cm

$^*$Corresponding author: V.I. Yukalov (E-mail: yukalov@theor.jinr.ru)

\newpage

\section{Introduction}

Magnetic nanoclusters are the objects enjoying rich and interesting properties
with a variety of applications, as can be inferred form the review articles
\cite{1,2,3,4,5}. For example, they are used in magnetic chemistry as catalysts;
in biomedical imaging for magnetic reading by magnetometers that measure the
magnetic field change induced by nanoclusters; in medical treatment by employing
alternating magnetic fields forcing the oscillation of the nanocluster magnetic
moment that produces heat destroying ill cells or bacteria; in genetic
engineering by attaching a nanocluster to a particular piece of the molecule
and then removing it together with this piece; in waste cleaning by stacking
nanoclusters to waste and then removing them together with waste; in information
storage, processing, and quantum computing; and so on \cite{1,2,3,4,5}. This is
why the methods of governing magnetization dynamics of nanoclusters are so much
important.

The use of nano sizes of clusters is principal, since such clusters behave as
large monodomain particles with a given large total spin. Typical radii of
nanoclusters are between 1 and 100 nm, containing from 100 to $10^5$ atoms.
Respectively, the effective total magnetization of a cluster can be of order
100 to $10^5 \mu_B$. The maximal size of a cluster, when it is in a monodomain
state, such that the spins of atoms, composing the cluster, sum up coherently,
forming an effective total spin, is characterized by the {\it coherence radius}.
The clusters, whose sizes are larger than the coherence radius, decompose into
several domains with oppositely directed magnetization, so that the total
cluster spin is zero.

Employing nanoclusters for the purpose of information storage, processing,
and quantum computing meets the requirements that contradict each other. From
one side, for information storage, it is required that the cluster magnetization
could be well frozen in a given direction, which needs the existence of a
sufficiently large magnetic anisotropy. But, from another side, for
information processing, it is necessary to be able to quickly reverse the
cluster magnetization, which is hindered by this anisotropy.

Nanoclusters enjoy the property to get their magnetization frozen at
temperatures lower than the blocking temperature that is of order of 10 - 100 K.
Then to reverse the cluster magnetization, one has to overcome the magnetic
anisotropy. So, for information storage, one needs large magnetic anisotropy,
while for information processing, the anisotropy has to be suppressed to
achieve fast magnetization reversal. To accomplish such a reversal, one applies
external transverse fields pushing the cluster magnetization. The discussion
of different methods of magnetization reversal by means of external magnetic
fields can be found in Refs. \cite{2,3,4,5,6}.

A very efficient method for realizing fast magnetization reversal of
nanoclusters is based on the use of feedback field from a resonant electric
circuit coupled to the ensemble of nanoclusters \cite{6,7,8,9,10}. Actually,
the idea that a resonant electric circuit can drastically shorten the
relaxation time of a spin system was advanced by Purcell \cite{11}, who
illustrated it for an ensemble of nuclear spins. This mechanism has later been
considered by Blombergen and Pound \cite{12}. The influence of the coupling
of a magnetic sample with a resonant circuit on magnetization dynamics is
called the {\it Purcell effect}. One sometimes calls it the radiation damping,
following Bloembergen and Pound \cite{12}. However the latter term is rather
a misnomer, as has been stressed by a number of researchers \cite{13,14,15},
since the coupling of a spin system with a resonator does strongly influences
spin dynamics, but not necessarily damping it. For instance, this effect
enhances nuclear magnetic resonance signals, which is widely used in NMR
techniques \cite{16,17,18}. The term "radiation damping" is also confusing
because of making impression that these are spins themselves that cooperate
by interacting with each other through a common radiation field. Such a
radiation correlation between radiating atoms is the essence of the Dicke
effect \cite{19}. But for magnetic particles, as has already been stressed
by Purcell \cite{11}, such a radiation collectivization of spin motion is
absolutely negligible. In atomic physics, one clearly distinguishes these
two principally different physical effects. And the term {\it Purcell effect}
is employed for describing the influence of a cavity resonator on atomic
radiation, which is the main part of the {\it cavity quantum electrodynamics}
\cite{20,21,22}.

The role of the Purcell effect on the spin dynamics of nuclear spins is well
studied. In resonance experiments, it enhances the NMR signals \cite{16,17,18}.
For strongly nonequilibrium systems of polarized nuclei, it leads to fast
magnetization reversal \cite{23,24,25,26,27} that has been discovered in
experiments \cite{23} and later confirmed in other experimental studies
(see references in the review article \cite{4}).

Magnetization dynamics in an ensemble of nanoclusters is essentially different
from that of nuclear systems. The basic differences are as follows.

(i) Nanoclusters possess rather large magnetic anisotropy that hinders the
possibility of simple regulation of spin motion.

(ii) Having large effective spins, nanoclusters also have strong spin dipole
interactions, which results in short dephasing time.

(iii) The most important difficulty for organizing collective spin motion in
a system of nanoclusters is the problem of cluster inhomogeneity, since
nanoclusters, being prepared by any of the known methods, whether by thermal
decomposition, or microemulsion reactions, or by thermal spraying, are not
identical particles, as nuclei would be. But nanoclusters differ by their
shapes and sizes, which results in the difference in the value of their
spins, dipole interactions, and of their anisotropies.

It is the aim of the present paper to investigate the magnetization dynamics
in a realistic inhomogeneous ensemble of nanoclusters exhibiting all their
typical properties of anisotropy and dipole interactions. Since analytic
investigation of such an inhomogeneous system is too much complicated, we
resort to computer modelling. Because the problem of magnetization reversal is
of special interest for many applications, such as information recording and
processing, we pay the main attention to finding the optimal conditions, when
the magnetization reversal is fast and as close to complete as possible.

\section{Realistic model of nanocluster system}

The sample formed by a system of nanoclusters is described by the Hamiltonian
\be
\label{1}
\hat H = \sum_i \hat H_i + \frac{1}{2} \sum_{i\neq j} \hat H_{ij} \; ,
\ee
where the first term characterizes single nanoclusters and the second term
describes their dipole interactions.

The typical Hamiltonian of a nanocluster has the form
\be
\label{2}
\hat H_i = - \mu_i {\bf B} \cdot {\bf S}_i - D(S_i^z)^2 + D_2 (S_i^y)^2 +
D_4 \left [(S_i^x)^2 (S_i^y)^2 + (S_i^y)^2 (S_i^z)^2 +
(S_i^z)^2 (S_i^x)^2 \right ] \; ,
\ee
in which the first term is the Zeeman energy, and other terms describe the
energy due to magnetic anisotropy, with the corresponding anisotropy
parameters \cite{1,2,3,4,5,6}.

The interaction Hamiltonian is caused by dipole spin interactions
\be
\label{3}
\hat H_{ij} = \sum_{\al\bt} D_{ij}^{\al\bt} S_i^\al S_j^\bt \; ,
\ee
with the dipolar tensor
\be
\label{4}
D_{ij}^{\al\bt} = \frac{\mu_0^2}{r_{ij}^3} \left ( \dlt_{\al\bt}
- 3 n_{ij}^\al n_{ij}^\bt \right ) \; ,
\ee
where
$$
r_{ij}\equiv |{\bf r}_{ij}| \; , \qquad
{\bf n}_{ij}\equiv \frac{{\bf r}_{ij}}{r_{ij}} \; , \qquad
{\bf r}_{ij} \equiv {\bf r}_i - {\bf r}_j \; .
$$

The total magnetic field, acting on the system, is the sum
\be
\label{5}
{\bf B} = B_0 {\bf e}_z + H{\bf e}_x
\ee
of an external constant field $B_0$ and the resonator feedback field $H$.

The considered sample is inserted into a coil of an electric circuit with an
attenuation $\gamma$ and natural frequency $\omega$. Moving spins of the
sample create electric current in the coil that, in turn, produces the feedback
magnetic field acting on these spins. The equation for the feedback field
follows from the Kirchhoff equation and can be written \cite{4,26,27} as
\be
\label{6}
\frac{dH}{dt} + 2\gm H + \om^2 \int_0^t H(t')\; dt' =
- 4\pi\eta\; \frac{dm_x}{dt} \; ,
\ee
where $\eta$ is filling factor and the electromotive force is caused by the
moving average magnetization density
\be
\label{7}
m_x = \frac{\mu_0}{V} \sum_j \lgl S_j^x \rgl \; .
\ee

The equations of spin dynamics are obtained in the following traditional way
\cite{4,14,28,29}. First, we write the Heisenberg equations of motion for the
spin components $S_j^\alpha$, in which the index $j = 1, 2, \ldots, N$
enumerates nanoclusters and $\alpha = x, y, z$. Then these equations are
averaged using semiclassical approximation, and spin attenuation is taken
into account by means of the second-order perturbation theory, as is described
in full details by Abragam and Goldman \cite{14,29}. This procedure
results in the equations that can effectively be obtained from the evolution
equations
$$
i\hbar \; \frac{d}{dt} \; S_j^x = [ S_j^x, H] - i \Gm_2 S_j^x \; ,
$$

$$
i\hbar \; \frac{d}{dt} \; S_j^y = [ S_j^y, H] - i \Gm_2 S_j^y \; ,
$$

\be
\label{8}
i\hbar \; \frac{d}{dt} \; S_j^z = [ S_j^z ,H] - i \gm_1
\left ( S_j^z - S \right ) \; ,
\ee
with the straightforward use of the semiclassical approximation \cite{4,14,28,29}.
Here $\Gamma_2$ is a transverse attenuation parameter, $\gamma_1$ is
longitudinal attenuation parameter, and $S \equiv \sum_{j=1}^N S_j/N$
is average spin value. In the case of strong initial polarization, the transverse
attenuation is characterized \cite{4,14,27,29} by the expression
\be
\label{9}
\Gm_2 = \gm_2 ( 1 - s^2 ) \; ,
\ee
where $s$ is the reduced average spin
\be
\label{10}
s \equiv \frac{1}{N} \sum_{j=1}^N \frac{\lgl S_j^z\rgl}{S_j} \; ,
\ee
and the natural width is
\be
\label{11}
 \gm_2 \equiv \frac{\rho \mu_0^2 S}{\hbar} =
\rho\hbar\gamma_S^2 S\;  .
\ee

In the following numerical calculations, we shall use the reduced
dimensionless attenuation parameters, measured in units of the Zeeman
frequency
\be
\label{12}
 \om_0 \equiv - \; \frac{\mu_0 B_0}{\hbar} = |\gamma_S|B_0 \; , \qquad
(\mu_0 = - 2 \mu_B =-\hbar|\gamma_S|) \;  .
\ee

Our goal is to investigate spin dynamics of nanoclusters with realistic
parameters. Thus, the values of the anisotropy parameters, typical for
nanoclusters, such as Co, Fe, and Ni nanoclusters, are
\be
\label{13}
 \frac{D}{\hbar\gm_2} = 10^{-3} \; , \qquad
 \frac{D_2}{\hbar\gm_2} = 10^{-3} \; , \qquad
\frac{D_4}{\hbar\gm_2} = 10^{-10} \; .
\ee

Studying spin dynamics, we shall take into account that real nanoclusters
are not completely identical with each other, but exhibit the dispersion
in their anisotropy parameters and spin values. The main aim of our study
is to analyze how the parameter dispersion influences spin dynamics and to
find conditions allowing for effective spin reversal.

\section{Introduction of dimensionless system parameters}

For numerical analysis, it is convenient to pass to dimensionless quantities.
To this end, we define the dimensionless resonator feedback field
\be
\label{14}
  p_H \equiv \frac{H}{B_0}
\ee
and the reduced transverse attenuation, caused by dipole interactions,
\be
\label{15}
 p_d \equiv \frac{\gm_2}{\om_0}\;   .
\ee
Also, we introduce the dimensionless anisotropy parameters
\be
\label{16}
p_A \equiv \frac{\om_A}{\om_0}\; , \qquad
p_B \equiv \frac{\om_B}{\om_0}\; , \qquad
p_C \equiv \frac{\om_C}{\om_0}\; ,
\ee
expressed through the anisotropy frequencies
\be
\label{17}
\om_A \equiv 2S \; \frac{D}{\hbar} \; , \qquad
\om_B \equiv 2S \; \frac{D_2}{\hbar} \; , \qquad
\om_C \equiv 2S^3 \; \frac{D_4}{\hbar} \;   .
\ee

In the equations of motion, we employ the reduced spin, with the components
\be
\label{18}
e_\al \equiv \frac{1}{N} \; \sum_{j=1}^N \; \frac{S_j^\al}{S_j} \; ,
\qquad e_z \equiv s \;   .
\ee
Since the nanocluster spins $S_j$ are large, of order $10^3$, the
semiclassical approximation is well justified. This allows us to treat
$\bf{e}$ as a classical vector. The evolution will be considered with respect
to the dimensionless time
\be
\label{19}
 \widetilde t \equiv \om_0 t \;  .
\ee

At the initial moment of time, there is no feedback field, which assumes the
initial conditions
\be
\label{20}
 p_H(0) = 0 \; , \qquad \dot{p}_H(0) = 0 \;  ,
\ee
where the overdot implies time derivative.

The initial conditions for the spin polarization $e_z(0)$ are constructed
as follows \cite{30,31}. For the given value of an initial polarization
$e_z(0)$, a variety of the initial orientations for the individual vectors
$S_j^z(0)$ are admissible. These orientations can be prescribed by a kind
of Monte Carlo techniques. First, a random configuration of vectors $S_j^z$
is taken and the corresponding total polarization is evaluated. A new
direction is chosen randomly for each spin, and the new total polarization
is calculated. If it is less than the initial one, the array with the
changed magnetic moment direction distribution is chosen as the second
iteration, otherwise, it is rejected. This procedure is repeated until the
system achieves the required average polarization $e_z(0)$ playing the role
of the initial condition. The initial spin polarization is assumed to be
directed opposite to the external magnetic field $B_0$.

\section{Optimal conditions for spin reversal}

We accomplish numerical solution of the evolution equations for
$N = 15^3 = 3375$ spins, searching for the conditions of the optimal spin
reversal. The Zeeman frequency $\omega_0$ is taken to be in resonance with
the circuit natural frequency $\omega$.

The peculiarity of spin reversal depends on the system parameters. For
instance, the optimal value of the resonator attenuation $\gamma$, defined
by the resonator circuit quality, essentially depends on the given
transverse attenuation $\gamma_2$, caused by spin dipole interactions.

Let us fix $\gamma_2 = 0.01 \omega_0$, which yields $p_d = 0.01$, and let
us take the anisotropy parameters typical for Co, Fe, and Ni nanoclusters,
as defined above. Below the blocking temperature, the longitudinal relaxation
is suppressed, so that $\gamma_1$ is much smaller than $\gamma_2$, which
allows us to set $\gamma_1/\gamma_2 = 10^{-3}$. The coil filling factor
is assumed to be close to one. The nanocluster spins are randomly distributed
by the normal law with mean $S = 1200$ and variance $\delta S = 400$.
Dynamics of spin polarization, $e_z$, is shown in Fig. 1 for different resonator
attenuations $\gamma$. Varying $\gamma$, we are looking for its optimal
value that provides the maximal permanent reversal, without superimposed
oscillations. It is this regime that is optimal for fast and stable
information processing \cite{32}. As is seen, for the given ratio
$\gamma_2/\omega_0$, the optimal value of the resonator attenuation is
$\gamma \approx 0.35\om_0$. Below this value, the reversal is slower and the
reversed spin value is smaller. For very low $\gamma$, there is no reversal
at all. Above the optimal value of $\gamma$, there arise oscillations,
and the reversed spin value diminishes.

The spin reversal is caused by the resonator feedback field. This is
illustrated in Fig. 2, where it is seen that the reversal occurs when
the amplitude of the feedback-field oscillations is maximal and almost
reaches the strength of the external field $B_0$. For each given ratio
$p_d = \gamma_2/\omega_0$, the corresponding optimal value of $\gamma$
is chosen. The stronger external field $B_0$, that is, the larger
$\omega_0$, hence, smaller $p_d$, the more effective is the reversal,
though the reversal time increases. Spin magnitudes are randomly
distributed, as in Fig. 1.

The dependence of the optimal $\gamma^*$ and the related value of the
reversed spin $s^* = e_z^*$, as functions of the ratio
$p_d = \gamma_2/\omega_0$, are presented in Fig. 3. The larger the
ratio $p_d$, the larger the optimal $\gamma^*$, but the smaller the value
of the reversed spin $s^*$.

In addition to spin dispersion, there can exist the dispersion of the
anisotropy parameters. We study spin dynamics, when the spins as well
as all anisotropy parameters are distributed by the normal law. The
spin mean is taken as $S = 1200$ and variance as $\delta S = 960$. The
mean values of the anisotropy parameters are assumed to be typical for
the Co, Fe, and Ni nanoclusters, as in Eq. (13), with the relative
variance
$$
 \frac{\dlt D}{D} = \frac{\dlt D_2}{D_2} =
\frac{\dlt D_4}{D_4} = 0.1 \;  .
$$
We accomplish 30 realizations of spin dynamics varying the anisotropy
parameters, with random spin distribution in each of the realizations.
The results, averaged over the realizations, are shown in Fig. 4.
We see that a rather strong dispersion of the system parameters does
not preclude spin reversal.

To understand when the spin reversal could be blocked by anisotropy,
we vary the parameters of the latter in a wide range, looking for the
values at which the reversal becomes blocked. Varying one of the
anisotropy parameters, we keep others fixed. Spins are randomly
distributed with mean $S = 1200$ and relative variance 0.3. Fig. 5
illustrates the results, from which the blocking anisotropy parameters
are evaluated as
$$
\frac{D}{\hbar \gm_2} \sim 10^{-2} \; , \qquad
\frac{D_2}{\hbar \gm_2} \sim 0.5\times 10^{-1} \; , \qquad
\frac{D_4}{\hbar \gm_2} \sim 10^{-8} \;  .
$$
These blocking parameters are essentially larger than the typical
values in Eq. (13). Hence, the typical anisotropy does not block
spin reversal.

With increasing external field $B_0$, that is, diminishing the ratio
$p_d = \gamma_2/\omega_0$, the blocking anisotropy parameters decrease.
This is illustrated by Fig. 6 that is to be compared with Fig. 5. So,
increasing the external field suppresses the role of the anisotropy.

Finally, we study whether a strong spin dispersion can destroy the
coherent spin dynamics and influence spin reversal. For this purpose,
we consider a random distribution of spins, with mean $S = 1200$ and
very large relative variance $\delta S / S = 1$. The results of
numerical simulations are given in Fig. 7 for the ratio
$p_d = \gamma_2/\omega_0 = 0.01$, with the corresponding optimal
$\gamma/\omega_0 = 0.3$, and varying anisotropy parameters. The results
demonstrate that, for small anisotropy parameters, spin dispersion
plays practically no role. This role increases for larger values of the
anisotropy parameters. This fact finds straightforward explanation from
the structure of the spin equations of motion. When the anisotropy
parameters are much smaller than the Zeeman frequency, the spin rotation
is governed by the same $\omega_0$ depending only on the external field
$B_0$, but independent of the spin lengths. But when the anisotropy
parameters are sufficiently large, approaching their blocking values,
then the anisotropy disturbs the effective rotation frequencies, so that
different spins rotate with different frequencies. This is equivalent
to the appearance of inhomogeneous broadening. Fortunately, the typical
nanocluster anisotropy parameters are much lower than their blocking
values. So, for a sufficiently strong external field, all spins rotate
with almost the same frequency $\omega_0$ and their dispersion even
being quite large, does not much influence their motion. Then the
coherent spin dynamics can be realized, resulting in a fast spin reversal.

\section{Conclusion}

We have studied the magnetization dynamics in an ensemble of magnetic
nanoclusters, keeping in mind a realistic situation, when the nanoclusters
are characterized by a microscopic Hamiltonian, with well defined
parameters. Another peculiarity of realistic nanocluster ensembles is
the dispersion in the values of their spins and anisotropy parameters,
which we also take into account. We have accomplished a series of computer
simulations varying the system properties and searching for the conditions
when the magnetization reversal is optimal, in the sense of being fast,
maximal, and quasi-stationary, without oscillations. Such a process of
magnetization reversal is necessary in a variety of applications. For
instance, for realizing effective quantum information processing.

For numerical values, we have chosen the nanocluster parameters typical
for Co, Fe, and Ni nanoclusters. It turns out that if the external magnetic
field defines the Zeeman frequency that is essentially larger than the
effective anisotropy frequencies, then the dispersion of spin magnitudes
plays practically no role. This also concerns the dispersion in the values
of the anisotropy parameters. The external magnetic field, sufficient for
this effect is 1 T, which corresponds to the Zeeman frequency
$\omega_0 \sim 10^{11}$ Hz.

We have shown that the magnetization reversal is caused by the resonator
feedback field that is self-organized by moving spins. The value of the
feedback field, in the moment of spin reversal, almost reaches the value
of the external field. Fast magnetization reversal can be achieved solely
by moving spins themselves, without involving strong transverse magnetic
fields. For such nanoclusters as Co, Fe, and Ni, the reversal time can be
of order $10^{-11}$ s. This conclusion suggests a convenient mechanism for
the efficient manipulation of nanocluster magnetization, which is of high
importance for a variety of applications.

\vskip 2mm

{\bf Acknowledgment}

\vskip 2mm

The authors acknowledge financial support from the Russian Foundation for
Basic Research under the projects 10-02-96023, 11-02-00086, 11-07-96007, and
12-02-00897. One of the authors (V.I.Y.) is grateful to E.P. Yukalova for 
useful discussions.

\newpage

\begin{center}
{\Large{\bf Figure Captions}}
\end{center}

\vskip 2cm

{\bf Fig. 1}. Dynamics of the average spin polarization $e_z = s$, for the
ratio $\gamma_2/\omega_0 = 0.01$, as a function of dimensionless time,
for random spins with the mean $S = 1200$ and relative spin variance
$\delta S / S = 0.333$ for different values of the resonator attenuation:
(a) $\gamma/\omega_0 = 0$ (dashed-dotted line),
$\gamma/\omega_0 = 0.04$ (dashed line),
$\gamma/\omega_0 = 0.12$ (solid line);
(b) $\gamma/\omega_0 = 0.3$ (dashed-dotted line),
$\gamma/\omega_0 = 0.4$ (dashed line),
$\gamma/\omega_0 = 0.8$ (solid line).

\vskip 1cm
{\bf Fig. 2}. Average spin polarization $e_z = s$ (dashed line) and feedback
field $p_H$ (solid line) as functions of dimensionless time, for different
ratios $p_d = \gamma_2/\omega_0$, with the corresponding optimal $\gamma$:
(a) $p_d = 0.1, \gamma/\omega_0 = 0.7$;
(b) $p_d = 0.01, \gamma/\omega_0 = 0.3$;
(c) $p_d = 0.001, \gamma/\omega_0 = 0.1$.
Spins are randomly distributed as in Fig.1.

\vskip 1cm
{\bf Fig. 3}. Dependence of the optimal $\gamma^*$ and the related value of the
reversed spin $s^* = e_z^*$, as functions of the ratio
$p_d = \gamma_2/\omega_0$.

\vskip 1cm
{\bf Fig. 4}. Influence of the combined dispersion of spins and anisotropy
parameters on the dynamics of spin reversal for
$p_d = \gamma_2/\omega_0 = 0.01$, with the corresponding optimal
$\gamma/\omega_0 = 0.3$:
(a) spin polarization $s = e_z$ (dashed line) and feedback field
$p_H$ (solid line) as functions of the dimensionless time for random spins
with the relative spin variance $\delta S / S = 0.8$ and fixed anisotropy
parameters; (b) Spin polarization $s = e_z$, averaged over 30 realizations
of the anisotropy distribution with the relative variance 0.1, as a
function of the dimensionless time; (c) feedback field $p_H$ as a function
of the dimensionless time, averaged over 30 realizations of the anisotropy
distribution.

\vskip 1cm
{\bf Fig. 5}. Spin polarization $s = e_z$, as a function of dimensionless time,
for the ratio $p_d = \gamma_2/\omega_0 = 0.01$, with the corresponding
optimal $\gamma/\omega_0 = 0.3$, for varying anisotropy parameters:
(a) $p_A = 0.1$ (solid line), $p_A = 0.2$ (dashed line),
$p_A = 0.3$ (dashed-dotted line), $p_A = 0.4$ (dotted line);
(b) $p_B = 0.5$ (solid line), $p_B = 1$ (dashed line),
$p_B = 2$ (dashed-dotted line), $p_B = 3$ (dotted line);
(c) $p_C = 0.1$ (solid line), $p_C = 0.2$ (dashed line),
$p_C = 0.3$ (dashed-dotted line), $p_C = 0.4$ (dotted line). In each case,
spins are randomly distributed with mean 1200 and relative variance 0.3.

\vskip 1cm
{\bf Fig. 6}. Spin polarization $s = e_z$, as a function of dimensionless time,
for the ratio $p_d = \gamma_2/\omega_0 = 0.01$, with the corresponding
optimal $\gamma/\omega_0 = 0.3$, for varying anisotropy parameters:
(a) $p_A = 0.01$ (solid line), $p_A = 0.05$ (dashed line),
$p_A = 0.1$ (dashed-dotted line), $p_A = 0.2$ (dotted line);
(b) $p_B = 0.05$ (solid line), $p_B = 0.1$ (dashed line),
$p_B = 0.2$ (dashed-dotted line), $p_B = 0.5$ (dotted line);
(c) $p_C = 0.01$ (solid line), $p_C = 0.05$ (dashed line),
$p_C = 0.1$ (dashed-dotted line), $p_C = 0.2$ (dotted line). In each case,
spins are randomly distributed with mean 1200 and relative variance 0.3.

\vskip 1cm
{\bf Fig. 7}. Spin polarization $s = e_z$, as a function of dimensionless
time, for the ratio $p_d = \gamma_2/\omega_0 = 0.01$, with the related optimal
$\gamma/\omega_0 = 0.3$, for varying anisotropy parameters. Solid line
describes the case of strong spin dispersion, with mean spin $S = 1200$
and relative variance $\delta S / S = 1$. Dashed line corresponds to the
case without spin dispersion. The anisotropy parameters are: 
(a) $p_A = 0.01$, $p_B = 0.01$, $p_C = 0.001$, 
(b) $p_A = 0.05$, $p_B = 0.1$, $p_C = 0.05$, 
(c) $p_A = 0.1$, $p_B = 0.2$, $p_C = 0.1$, 
(d) $p_A = 0.2$, $p_B = 0.7$, $p_C = 0.2$, 
(e) $p_A = 0.05$, $p_B = 0.01$, $p_C = 0.001$, 
(f) $p_A = 0.01$, $p_B = 0.1$, $p_C = 0.001$,
(g) $p_A = 0.01$, $p_B = 0.01$, $p_C = 0.02$.

\newpage

%Figure 1
\begin{figure}[ht]
\vspace{9pt}
\centerline{\includegraphics[width=14cm]{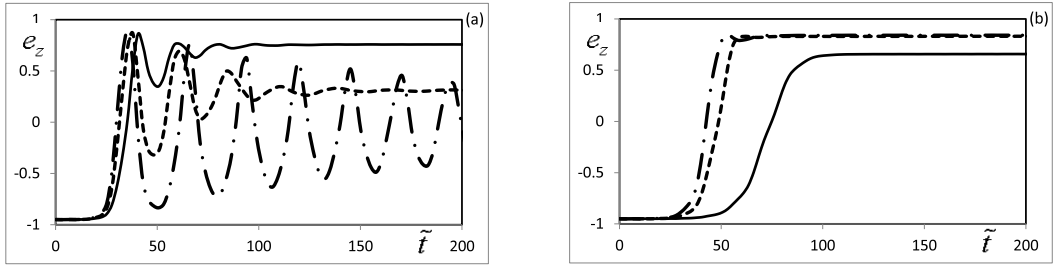} }
\caption{Dynamics of the average spin polarization $e_z = s$, for the
ratio $\gamma_2/\omega_0 = 0.01$, as a function of dimensionless time,
for random spins with the mean $S = 1200$ and relative spin variance
$\delta S / S = 0.333$ for different values of the resonator attenuation:
(a) $\gamma/\omega_0 = 0$ (dashed-dotted line),
$\gamma/\omega_0 = 0.04$ (dashed line),
$\gamma/\omega_0 = 0.12$ (solid line);
(b) $\gamma/\omega_0 = 0.3$ (dashed-dotted line),
$\gamma/\omega_0 = 0.4$ (dashed line),
$\gamma/\omega_0 = 0.8$ (solid line).}
\label{fig:Fig.1}
\end{figure}

\newpage

%Figure 2
\begin{figure}[ht]
\vspace{9pt}
\centerline{\includegraphics[width=14cm]{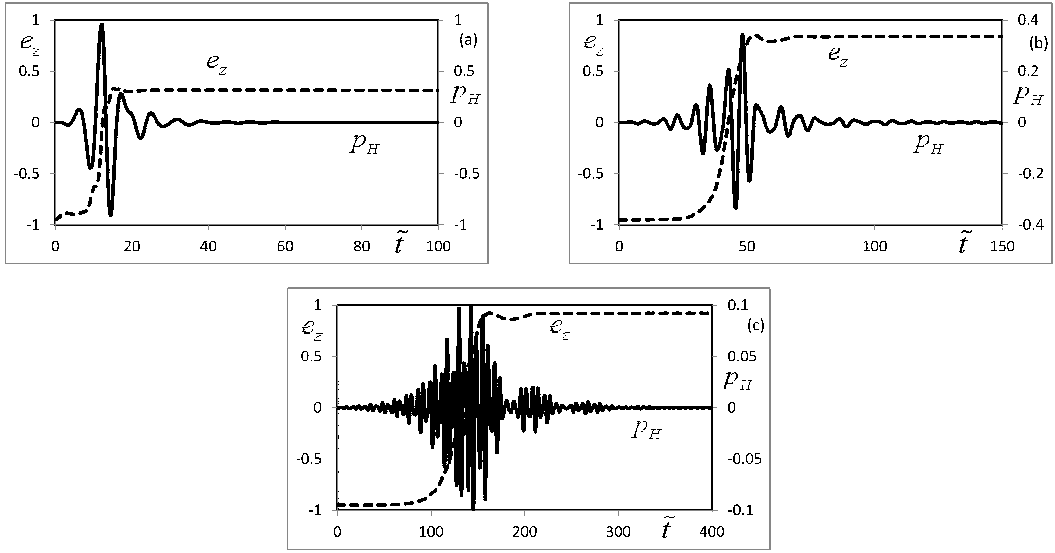} }
\caption{Average spin polarization $e_z = s$ (dashed line) and feedback
field $p_H$ (solid line) as functions of dimensionless time, for different
ratios $p_d = \gamma_2/\omega_0$, with the corresponding optimal $\gamma$:
(a) $p_d = 0.1, \gamma/\omega_0 = 0.7$;
(b) $p_d = 0.01, \gamma/\omega_0 = 0.3$;
(c) $p_d = 0.001, \gamma/\omega_0 = 0.1$.
Spins are randomly distributed as in Fig.1.}
\label{fig:Fig.2}
\end{figure}

\newpage

%Figure 3
\begin{figure}[ht]
\vspace{9pt}
\centerline{\includegraphics[width=14cm]{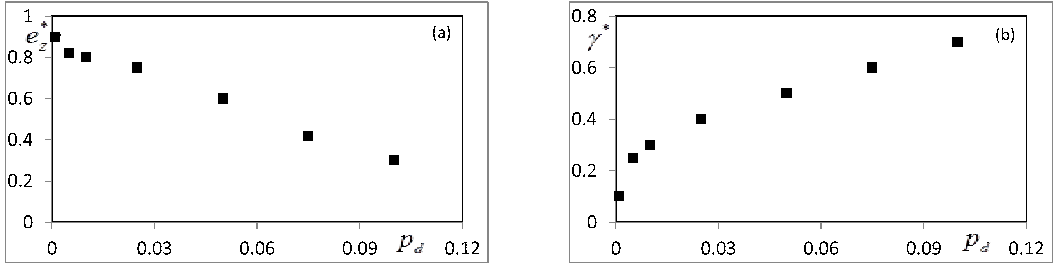}  }
\caption{Dependence of the optimal $\gamma^*$ and the related value 
of the reversed spin $s^* = e_z^*$, as functions of the ratio
$p_d = \gamma_2/\omega_0$.}
\label{fig:Fig.3}
\end{figure}

\newpage

%Figure 4
\begin{figure}[ht]
\vspace{9pt}
\centerline{\includegraphics[width=14cm]{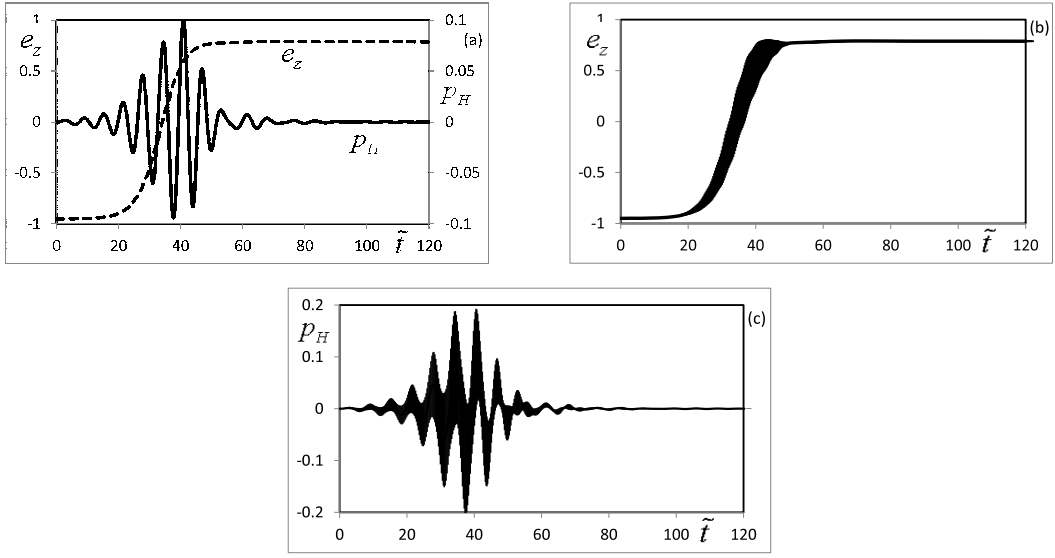}}
\caption{Influence of the combined dispersion of spins and anisotropy
parameters on the dynamics of spin reversal for $p_d=\gamma_2/\omega_0=0.01$,
with the corresponding optimal $\gamma/\omega_0 = 0.3$:
(a) spin polarization $s = e_z$ (dashed line) and feedback field
$p_H$ (solid line) as functions of the dimensionless time for random 
spins with the relative spin variance $\delta S / S = 0.8$ and fixed 
anisotropy parameters; (b) Spin polarization $s = e_z$, averaged over 
30 realizations of the anisotropy distribution with the relative 
variance 0.1, as a function of the dimensionless time; (c) feedback 
field $p_H$ as a function of the dimensionless time, averaged over 
30 realizations of the anisotropy distribution.}
\label{fig:Fig.4}
\end{figure}

\newpage

%Figure 5
\begin{figure}[ht]
\vspace{9pt}
\centerline{\includegraphics[width=14cm]{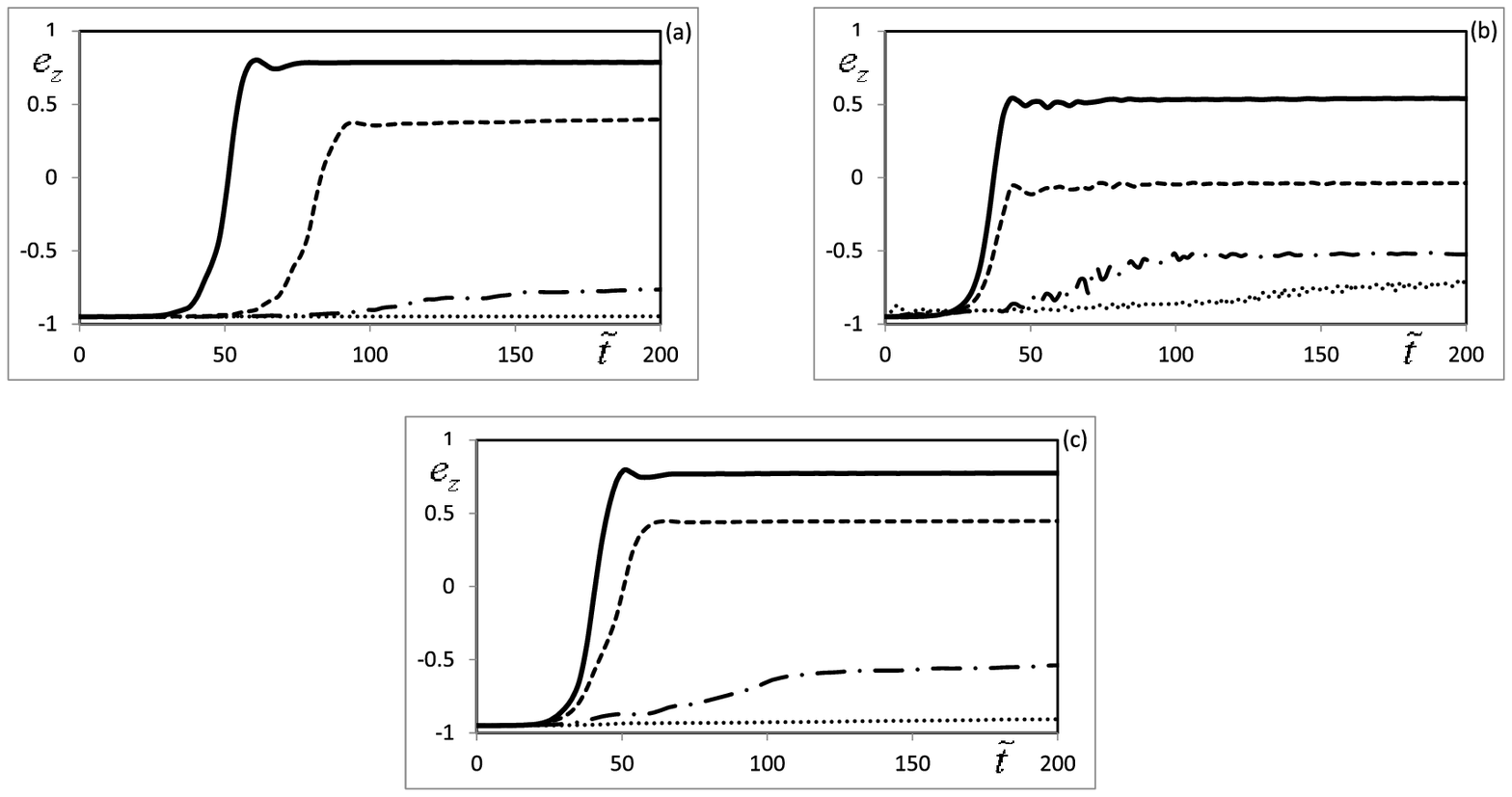} }
\caption{Spin polarization $s = e_z$, as a function of dimensionless time,
for the ratio $p_d = \gamma_2/\omega_0 = 0.01$, with the corresponding
optimal $\gamma/\omega_0 = 0.3$, for varying anisotropy parameters:
(a) $p_A = 0.1$ (solid line), $p_A = 0.2$ (dashed line), $p_A = 0.3$ 
(dashed-dotted line), $p_A = 0.4$ (dotted line); (b) $p_B = 0.5$ (solid line),
$p_B = 1$ (dashed line), $p_B = 2$ (dashed-dotted line), $p_B = 3$ (dotted 
line); (c) $p_C = 0.1$ (solid line), $p_C = 0.2$ (dashed line), $p_C = 0.3$
(dashed-dotted line), $p_C = 0.4$ (dotted line). In each case, spins are 
randomly distributed with mean 1200 and relative variance 0.3.}
\label{fig:Fig.5}
\end{figure}

\newpage

%Figure 6
\begin{figure}[ht]
\vspace{9pt}
\centerline{\includegraphics[width=14cm]{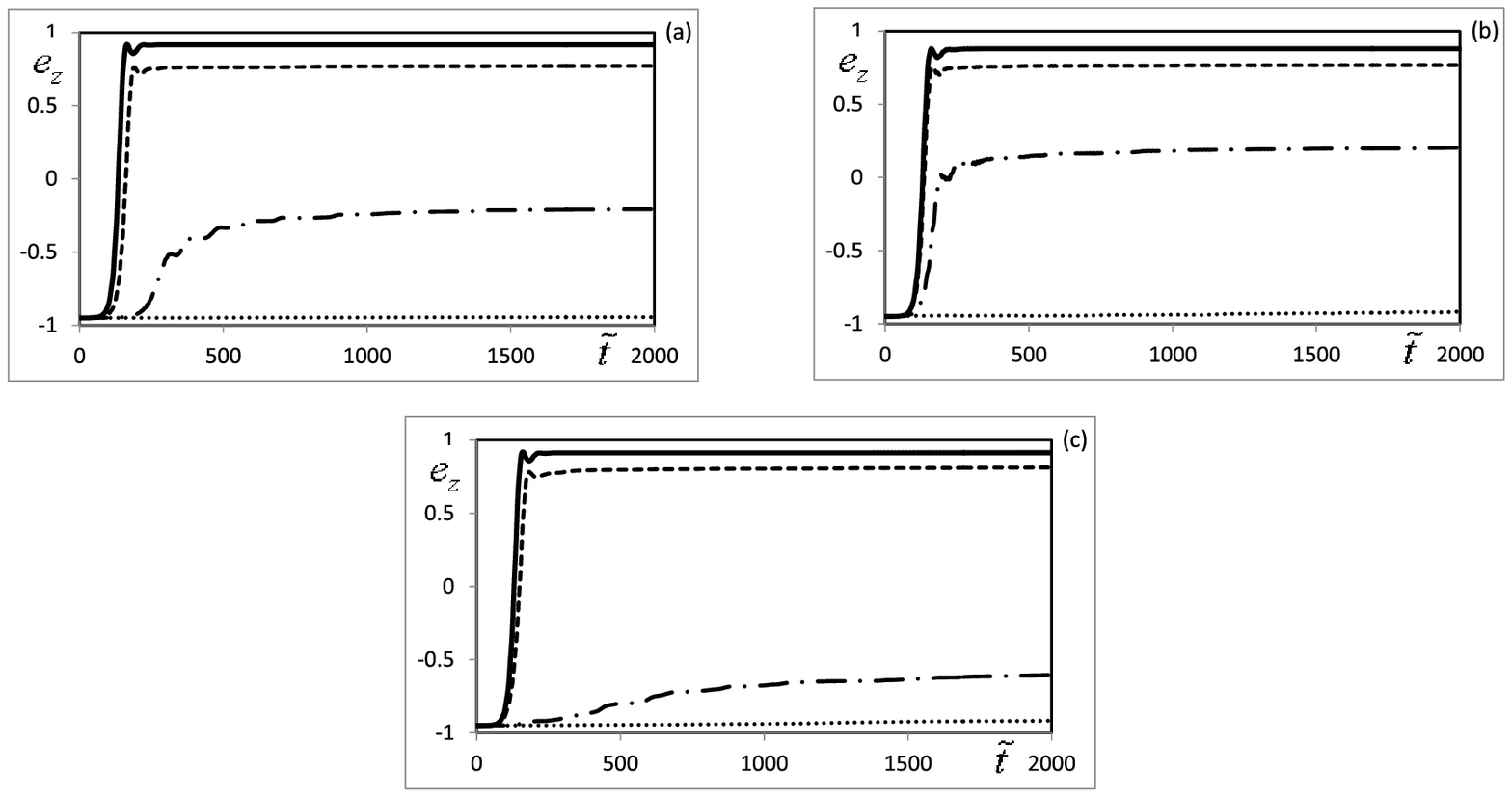}  }
\caption{Spin polarization $s = e_z$, as a function of dimensionless time,
for the ratio $p_d = \gamma_2/\omega_0 = 0.01$, with the corresponding
optimal $\gamma/\omega_0 = 0.3$, for varying anisotropy parameters:
(a) $p_A = 0.01$ (solid line), $p_A = 0.05$ (dashed line), $p_A = 0.1$ 
(dashed-dotted line), $p_A = 0.2$ (dotted line); (b) $p_B = 0.05$ (solid 
line), $p_B=0.1$ (dashed line), $p_B=0.2$ (dashed-dotted line), $p_B=0.5$ 
(dotted line); (c) $p_C = 0.01$ (solid line), $p_C = 0.05$ (dashed line),
$p_C = 0.1$ (dashed-dotted line), $p_C = 0.2$ (dotted line). In each case,
spins are randomly distributed with mean 1200 and relative variance 0.3.}
\label{fig:Fig.6}
\end{figure}

\newpage

%Figure 7
\begin{figure}[ht]
\vspace{9pt}
\centerline{\includegraphics[width=14cm]{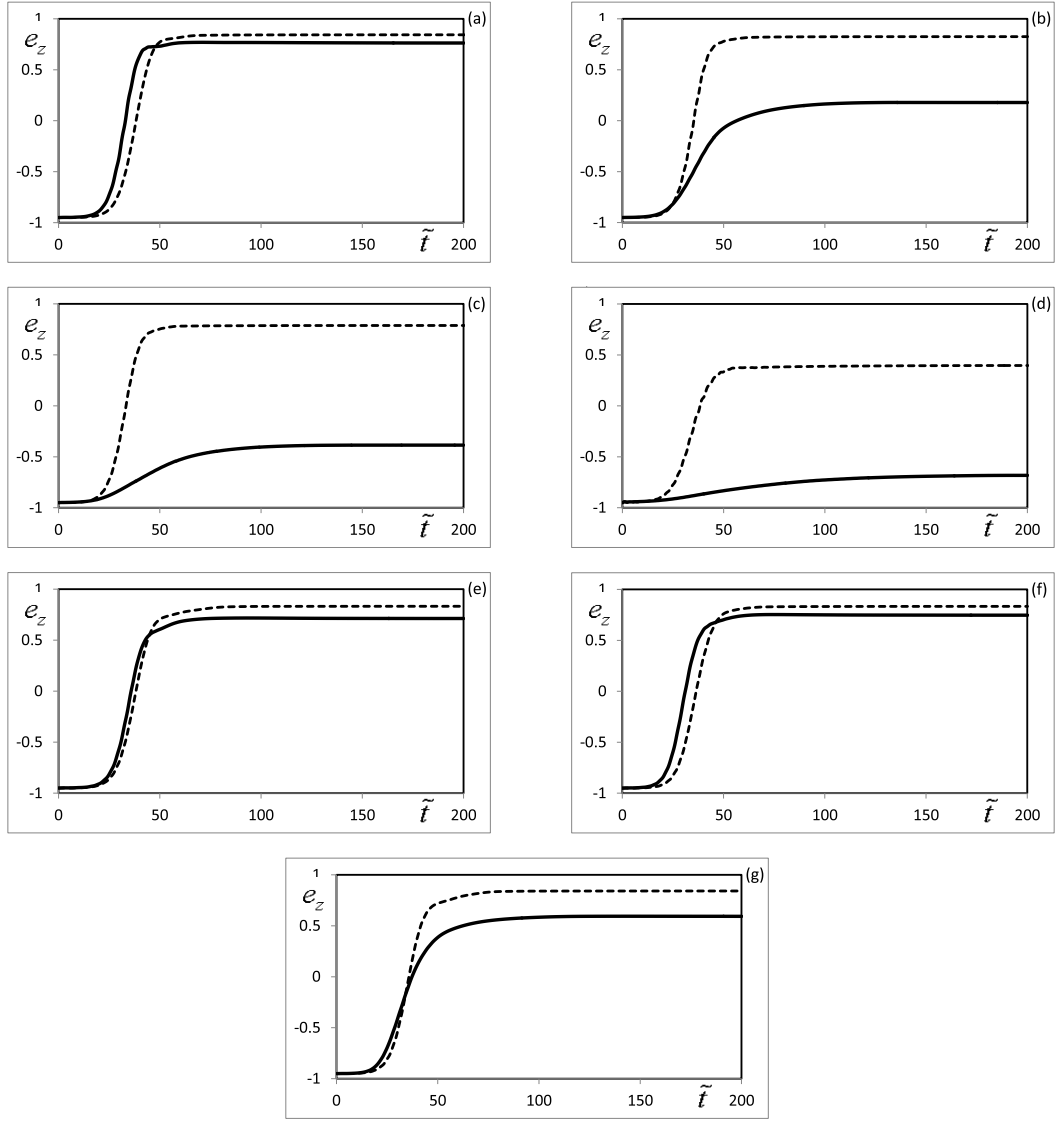} }
\caption{Spin polarization $s = e_z$, as a function of dimensionless
time, for the ratio $p_d = \gamma_2/\omega_0 = 0.01$, with the related optimal
$\gamma/\omega_0 = 0.3$, for varying anisotropy parameters. Solid line
describes the case of strong spin dispersion, with mean spin $S = 1200$
and relative variance $\delta S / S = 1$. Dashed line corresponds to the
case without spin dispersion. The anisotropy parameters are: 
(a) $p_A = 0.01$, $p_B = 0.01$, $p_C = 0.001$, 
(b) $p_A = 0.05$, $p_B = 0.1$, $p_C = 0.05$, 
(c) $p_A = 0.1$, $p_B = 0.2$, $p_C = 0.1$, 
(d) $p_A = 0.2$, $p_B = 0.7$, $p_C = 0.2$, 
(e) $p_A = 0.05$, $p_B = 0.01$, $p_C = 0.001$, 
(f) $p_A = 0.01$, $p_B = 0.1$, $p_C = 0.001$,
(g) $p_A = 0.01$, $p_B = 0.01$, $p_C = 0.02$.}
\label{fig:Fig.7}
\end{figure}

\end{document}